\documentclass[a4paper,12pt]{article}
\usepackage[left=3cm,top=2cm,right=1cm,bottom=2cm,nohead]{geometry}
\usepackage[colorlinks=false,pdfborder={0 0 0}]{hyperref}
\usepackage{cite}
\usepackage{graphicx}

\title{Herding interactions as an opportunity to prevent extreme events in financial markets}
\author{Aleksejus Kononovicius\thanks{email: \href{mailto:aleksejus.kononovicius@tfai.vu.lt}{aleksejus.kononovicius@tfai.vu.lt}; website: \url{http://kononovicius.lt}}, Vygintas Gontis \thanks{email: \href{mailto:vygintas@gontis.eu}{vygintas@gontis.eu}; website: \url{http://gontis.eu}}}
\date{}

\newcommand{\rmd}{\mathrm{d}}

\begin{document}

\maketitle

\begin{abstract}
A characteristic feature of complex systems in general is a tight coupling between their
constituent parts. In complex socio-economic systems this kind of behavior leads to self-organization,
which may be both desirable (e.g. social cooperation) and undesirable (e.g. mass panic, financial
``bubbles'' or ``crashes''). Abundance of the empirical data as well as general insights
into the trading behavior enables the creation of simple agent-based models reproducing
sophisticated statistical features of the financial markets.
In this contribution we consider a possibility to prevent self-organized extreme events in financial
market modeling its behavior using agent-based herding model, which reproduces main stylized facts of the financial markets. We show that introduction of agents with
predefined fundamentalist trading behavior helps to significantly reduce the probability of the extreme
price fluctuations events. We also investigate random trading, which was previously found to
be promising extreme event prevention strategy, and find that its impact on the market has to be considered among other opportunities to stabilize the markets.
\end{abstract}

\section{Introduction}

Empirical data from complex socio-economic systems is known to frequently follow power law distributions \cite{Bouchaud2004Cambridge,Chakraborti2011RQUF1,Cont1997Springer,Gabaix2009AR,Gontis2012ACS,
Karsai2012NIH,Kondor2014PlosOne,Mantegna2000Cambridge,Schinckus2013ConPhys}.
This basically means that the extreme events in such systems become significantly more probable and consequently more frequent.
In certain cases these extreme events may be desirable - one may see emergence of ``homo socialis'', social cooperation, fads
and norms as the most straightforward examples \cite{Challet1997PhysA,Galam2007PhysA,Galam2010PhysA,Helbing2012Springer,
Schweitzer2013ACS}. But there are also undesirable extreme events - such as mass panic, financial ``bubbles'' and ``crashes''.
Either way these events are thought to be caused by the same social interaction mechanism - herding behavior,
which encourages endogenous self-organization of the socio-economic systems \cite{Akerlof2009Princeton,Becker1991JPolitEco,
Biondo2013PhysRevE,Bouchaud2013JStatPhys,Cajueiro2009CSF,Galam2010PhysA,Dyer2009RSTB,Helbing2012Springer,
Schweitzer2013ACS,Parisi2013PhysRevE,Raafat2009TCogSci}. In this paper we address the possibility to control financial
market fluctuations, via herding interactions, thus preventing extreme events, or greatly reducing their probability.

Heterogeneous agent-based modeling \cite{Chakraborti2011RQUF2,Cincotti2008CompEco,Conte2012EPJ,Cristelli2012Fermi,
Farmer2009Nature,Frederick2013PNAS,Samanidou2007RepProgPhys,Yakovenko2009Springer} is the most suitable framework to test
the idea of using herding interactions to prevent extreme events. This framework introduces a generalized
concept of an agent, which is meant to be used in place of the interacting parts of the modeled system. Interactions between
those agents are also generalized and simplified. After this simplification only statistically relevant general behavioral details are
retained and thus the resulting models appear to be simple and able to capture important features of the modeled systems. Though unlike in mainstream economics, the agents and their interactions are not oversimplified - reduced to the representative agent. Collective interactions of heterogeneous agents are expected to cause emergent behavior, emerging both from a very simple ruleset \cite{Challet2000PhysRevLett,Farmer2005PNAS} as well as more realistic and complex setups \cite{Lux1999Nature,Preis2006EPL,Chiarella2008,Cristelli2014Thesis,JacobLeal2014}.

We base our research on a simple agent-based model proposed by Alan Kirman, see \cite{Kirman1993QJE},
which primarily explains the herding behavior in ant colonies. Various minimal modifications of this model were shown to be
able to reproduce the simplest stylized facts observed in the financial markets - such as power law distributions
\cite{Alfarano2005CompEco,Alfarano2008Dyncon,Alfarano2013EJF,Alfi2009EPJB1,Alfi2009EPJB2,Kononovicius2012PhysA} as
well as power law spectral density and auto-correlations \cite{Kirman2002SNDE,Kononovicius2012PhysA}. A
more sophisticated treatment of the agent-based herding model further enhances the quality of reproduced
statistical features - three state model was shown to reproduce fractured spectral density \cite{Kononovicius2013EPL}.
While the latest developments enable reproduction of the detailed empirical probability density function (abbr. PDF) and spectral density of high-frequency absolute return \cite{Gontis2014PlosOne}.

In the following section we will introduce controlled agents into the generic Kirman's model and discuss the impact of this predefined agent behavior. Next we will analyze three state artificial financial market built upon the generic herding model and explore the effectiveness of certain extreme event prevention strategies. We will consider straightforward one, based on the market fundamentals, and less predictable one, based on random trading (proposed by \cite{Biondo2013PhysRevE,Biondo2013PlosOne}). Finally we will conclude the paper by discussing obtained results.

\section{Agent-based herding model with controlled agents}
\label{sec:generic}

In numerous previous works \cite{Alfarano2005CompEco,Alfarano2008Dyncon,Alfarano2013EJF,Alfi2009EPJB1,
Alfi2009EPJB2,Kirman2002SNDE,Kononovicius2012PhysA,Kononovicius2013EPL,Gontis2014PlosOne} artificial agent-based financial markets were constructed using a very simple agent-based herding model, originally proposed by Alan Kirman in \cite{Kirman1993QJE}. In this section we will briefly discuss Kirman's agent-based herding model and will introduce two types of controlled agents, ones with predefined behavior, into the original model.

\begin{figure}
\centering
\includegraphics[width=12cm]{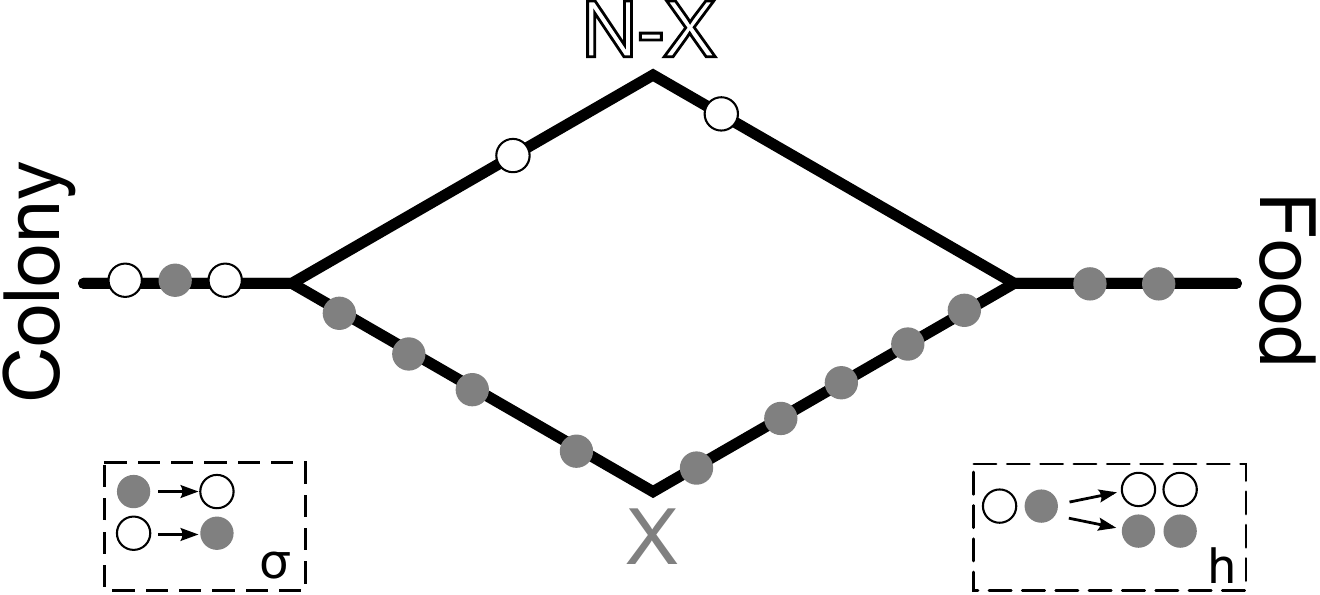}
\caption{Schematic representation of the experiment and mathematical herding model. Ants are moving from the colony (left hand side) to the food source (right hand side) using one of the two available paths. Switching between the paths may be based either on individual decision, single agent transitions, or be caused by exchange of information, two agent interactions.}
\label{fig:schema}
\end{figure}

Biological research \cite{Pastels1987Birkhauser1,Pastels1987Birkhauser2,Krause2011TEcoEvo}, as well as numerous socio-economic papers \cite{Becker1991JPolitEco,Dyer2009RSTB,Raafat2009TCogSci}, provides extensive evidence that herding interactions play crucial role in social scenarios both in animal and human societies. In the experimental setup by Pasteels \cite{Pastels1987Birkhauser1,Pastels1987Birkhauser2} ants were allowed to move from their colony to food source using one of the two available paths (see schematic representation of the experiment in Fig. \ref{fig:schema}). It would be far more convenient for them to use both paths at once as it would increase throughput, but at any given time majority of ants tend to use only a single path. Interestingly enough it was observed that from time to time this ``chosen'' path is switched purely due to endogenous interactions. So despite symmetrical setup asymmetric switching behavior was observed.

In order to describe such  behavior Kirman proposed \cite{Kirman1993QJE} a simple one-step transition model. In contemporary form the one-step transition probabilities of this model are expressed as follows \cite{Alfarano2005CompEco,Kononovicius2012PhysA}
\begin{eqnarray}
& p (X \rightarrow X+1) = (N-X) ( \sigma_1 + h X ) \Delta t , \\
& p (X \rightarrow X-1) = X [ \sigma_2 + h (N-X) ] \Delta t ,
\end{eqnarray}
here $N$ is a fixed number of agents in the system (thus one of the available states is occupied by $X$ agents and
the other by  $N-X$ agents), $\sigma_i$ represent idiosyncratic transition rates to state $i$, $h$ is the intensity
of herding behavior and $\Delta t$ is a short time step (during which only one transition would be probable).

In the above we have assumed that all agents may interact with all other agents, or namely they act on a global scale.
It is known that local interactions lead to the extensive statistics and may be described using ordinary differential equation.
While global interactions, as chosen for the financial market setup, lead to non-extensive statistics
\cite{Alfarano2009Dyncon,Kononovicius2014EPJB}
and are well described (for $x = X/N$ in the limit $N \rightarrow \infty$) by the following stochastic differential equation
\cite{Alfarano2005CompEco,Kononovicius2012PhysA}:
\begin{equation}
\rmd x = [ \sigma_1 (1-x) - \sigma_2 x ] \rmd t + \sqrt{2 h x (1-x)} \rmd W , \label{eq:nonextsde}
\end{equation}
here $W$ stands for a standard one dimensional Brownian motion (or alternatively - Wiener process). The stationary PDF of $x$, in this case, is given by
\begin{equation}
P_0(x) = \frac{\Gamma(\frac{\sigma_1}{h}+\frac{\sigma_2}{h})}{\Gamma(\frac{\sigma_1}{h}) \Gamma(\frac{\sigma_2}{h})} x^{\frac{\sigma_1}{h}-1} (1-x)^{\frac{\sigma_2}{h}-1} , \label{eq:nonextpdfx}
\end{equation}

The model can be now extended by including $M$ agents, which are controlled externally and switch states only due to this exogenous influence. For the sake of convenience we can assume that $M_1$ of them are in the state $1$, while $M_2$ are in the state $2$. Namely unlike the other agents, the controlled agents do not switch their state due to endogenous interactions, though they are allowed to trigger endogenous switches of the other agents. In this case one-step transition probabilities take the following form:
\begin{eqnarray}
& p (X \rightarrow X+1) = (N-X) [ \sigma_1 + h (X+M_1) ] \Delta t , \\
& p (X \rightarrow X-1) = X [ \sigma_2 + h (N-X+M_2) ] \Delta t ,
\end{eqnarray}
It is evident that one can include herding terms with $M_1$ and $M_2$ into spontaneous transition rates, $\sigma_i$. Namely, one can set $\tilde{\sigma}_1 = \sigma_1 + h M_1$
and $\tilde{\sigma}_2 = \sigma_2 + h M_2$  to get back to the original form of the herding model just with shifted individual preferences, $\tilde{\sigma}_i$. It means that Eqs. (\ref{eq:nonextsde}) and (\ref{eq:nonextpdfx}) are valid for the system with controlled agents, just $\sigma_1$ and $\sigma_2$ are replaced with $\tilde{\sigma}_1$ and $\tilde{\sigma}_2$.

If we have a system with $M$ controlled agents, which stochastically switch their state with equal probabilities, then we can assume that $M_1=M_2=M/2$. Such interpretation of controlled agents impact on the system is very valuable in the context of ongoing discussion about the role of random trading strategies in financial markets \cite{Biondo2013PhysRevE,Biondo2013PlosOne}. It opens up two distinct possibilities to control the behavior of the two state agent system - one can introduce a few agents with fixed states or just let them stochastically switch between available states. In both cases the standard deviation of stationary PDF of $x$ can be reduced.

\begin{figure}
\centering
\includegraphics[width=7cm]{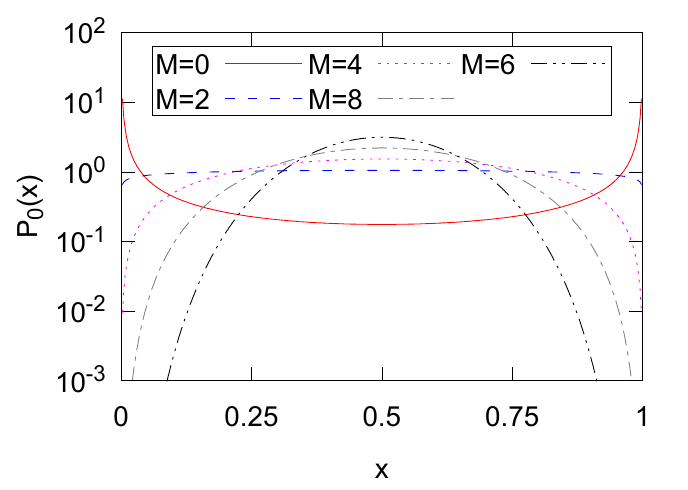}
\caption{Stationary PDF, $P_0(x)$, of agent population in the symmetric case of the two state herding model. Parameters were set as follows: $\tilde{\varepsilon} = 0.1 + M/2$, $M=0;2;4;8;16$. The corresponding parameters of a $q$-Gaussian distribution, Eq. (\ref{eq:nonextpdf}), are $q=-9; 0.09; 0.375; 0.677; 0.859$ and $\sigma_q = \frac{1-q}{3-q}$.}
\label{fig:analytical}
\end{figure}

In the symmetric case, $\tilde{\varepsilon} = \frac{\tilde{\sigma}_1}{h} = \frac{\tilde{\sigma}_2}{h}$, with $\tilde{\varepsilon}>1$ the stationary PDF, Eq. (\ref{eq:nonextpdfx}), may be rewritten as $q$-Gaussian (for more information on $q$-Gaussian and non-extensive statistics see \cite{GellMann2004Oxford})
\begin{equation}
P_0(x) = C_{q}\exp_{q} \left[-\frac{4}{1-q} \left( x-\frac{1}{2} \right)^2 \right] = \frac{(3-q) \Gamma(\frac{3-q}{2(1-q)})}{\sqrt{\pi} \Gamma(\frac{1}{1-q})} \left[4(x-x^2) \right]^{\frac{1}{1-q}} , \label{eq:nonextpdf}
\end{equation}
where $q=1-\frac{1}{\tilde{\varepsilon}-1} = 1 - \frac{2}{2 \varepsilon + M - 2}$. For $M>0$, having in mind that $M$ is positive even integer and that $\varepsilon >0$, entropic index, $q$, is smaller than $1$. Note that as $\varepsilon$ and $M$ increases $q$ approaches 1, and thus the distribution grows similar to the truncated Gaussian distribution. In Fig. \ref{fig:analytical} we demonstrate the stationary PDF, $P_0(x)$, calculated from the Eq. (\ref{eq:nonextpdf}) for several values of predefined agents $M$. From the theoretical point of view this herding model serves as a very simple example of a stochastic system on the macroscopic level, which can be stabilized by introducing agents, who act stochastically.

In the next section we follow up by considering a three state agent-based herding model of the financial markets. Introduction of the controlled agents in this setup is no longer trivial and few different approaches appear to be equally viable. Those different approaches lead to different results, which may impact the usability of extreme event prevention strategies.

\section{Extreme event prevention in a more sophisticated financial market model}

\label{sec:threestate}

In previous section we considered a possibility to stabilize macroscopic fluctuations of the two state herding model by introducing controlled agents. In the previous setup the impact of controlled agents is rather straightforward, see \cite{Kononovicius2014PhysA} for a more detailed consideration. In this section we consider the three state herding model of the financial markets, proposed in \cite{Kononovicius2013EPL,Gontis2014PlosOne}, introducing  stochastic and predefined agents into it.

\begin{figure}
\centering
\includegraphics[width=7cm]{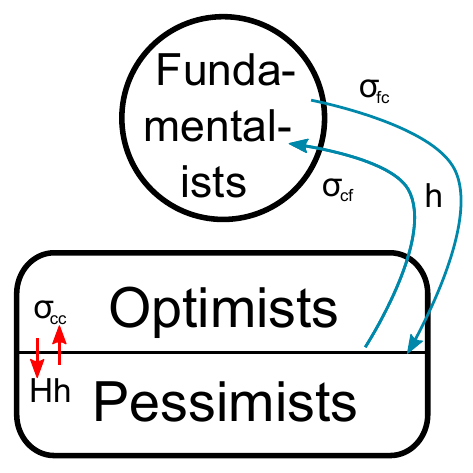}
\caption{Schematic representation of the three state herding model. Simplified model parameters $\sigma_{ij}$ describe idiosyncratic switching rate from state $i$ to state $j$, $h$ describes herding behavior between fundamentalists and chartists (optimists and pessimists collectively) and $H h$ describes herding behavior between optimists and pessimists.}
\label{fig:schema2}
\end{figure}

Let us now briefly discuss the basics of the three state herding model of financial markets. In this model the financial fluctuations are derived from the population dynamics of the three agent groups: fundamentalists (we will use subscript $f$ to denote them) and chartists (subscript $c$), who may be either optimists (subscript $o$) or pessimists (subscript $p$). The schematic representation of the three state herding model is given in Fig. \ref{fig:schema2}.

The main idea of this model is to link endogenous dynamics of the agents to the market price movements. Instantaneous occupations of the agent states impact the demand and supply in the artificial market, and thus determine the equilibrium price.
We define fundamentalists as agents aware of market fundamentals, which we assume to be quantified by the fundamental price, $P_f$. Namely, fundamentalists expect that market price, given enough time, will approach fundamental price. Having in mind our work reproducing statistical properties of real markets it is possible to treat fundamentalists as long time traders, which act on the longer time scales than chartists up to three orders of magnitude.
Chartists, on the other hand, simply switch between the buying (optimism) and selling (pessimism) behavior, switching their opinions quickly. One can view chartists as short term speculative traders, contributing to very rapid price movements.
These assumptions provide us with demands generated by those of agents groups:
\begin{eqnarray}
& D_f = X_f [ \ln P_f - \ln P(t) ] , \\
& D_c = r_0 (X_o - X_p) = r_0 X_c \xi ,
\end{eqnarray}
where $\xi = \frac{X_o - X_p}{X_c}$ is the average mood of chartists.
One can obtain price by applying Walrasian scenario \cite{Kononovicius2013EPL,Gontis2014PlosOne}, which results in:
\begin{equation}
p(t) = \ln \frac{P(t)}{P_f} = r_0 \frac{X_c}{X_f} \xi .
\end{equation}
In the above $r_0$ describes relative impact of chartist trader. The larger $r_0$, the bigger fluctuations of price occur, so $r_0$ can be folded into other empirically defined parameters. Further in this paper, for the sake of simplicity, we assume $r_0 =1$.

To simplify model we need to assume that there is no significant qualitative difference between optimists and pessimists (symmetry of idiosyncratic switching rates, folding of fundamentalist-optimist and fundamentalist-pessimist interactions into fundamentalist-chartist interaction) and also that chartist agents trade noticeably more frequently. After making these assumptions we may take the limit $N \rightarrow \infty$ to arrive to the following set of stochastic differential equations (detailed mathematical derivation may be found in \cite{Kononovicius2013EPL,Gontis2014PlosOne}):
\begin{eqnarray}
& \rmd x_f = \frac{(1-x_f) \varepsilon_{cf} - x_f \varepsilon_{fc}}{\tau(x_f,\xi)} \rmd t + \sqrt{\frac{2 x_f (1-x_f)}{\tau(x_f,\xi)}} \rmd W_{f} , \label{eq:nftau}\\
& \rmd \xi = - \frac{2 H \varepsilon_{cc} \xi}{\tau(x_f,\xi)} \rmd t + \sqrt{\frac{2 H (1-\xi^2)}{\tau(x_f,\xi)}} \rmd W_{\xi} . \label{eq:xitau}
\end{eqnarray}
Note that here we have introduced scaled time, $t_s = h t$, (though we omit subscript in the equations) and appropriately rescaled model parameters. Namely the idiosyncratic transition rates are now given by $\varepsilon_{cf} = \sigma_{cf} / h$, $\varepsilon_{fc} = \sigma_{fc} / h$ and $\varepsilon_{cc} = \sigma_{cc} / (H h)$, while $H$ gives us the relative intensity of chartist-chartist herding interactions  in respect to chartist-fundamentalist interactions. Note that Eqs. (\ref{eq:nftau}) and (\ref{eq:xitau}) are coupled only through the inter-event time $\tau(x_f,\xi)$. The inter-event time, $\tau(x_f,\xi)$, serves as a macroscopic state feedback on the microscopic activity of all agents,
\begin{equation}
\frac{1}{\tau(x_f,\xi)}= \left( 1 + a \left| r_0 \frac{1-x_f}{x_f} \xi \right| \right)^{\alpha} =\left( 1 + a \left| p(t) \right| \right)^{\alpha}. \label{eq:taunfxi}
\end{equation}
Numerous empirical analyses (e.g., \cite{Rak2013APP}) provide a background for the chosen relationship between inter-event time, $\tau(x_f,\xi)$, and relative log-price, $p(t)$, as well as suggest that the value of exponent, $\alpha = 2$. Note that there is a difference in Eq. (\ref{eq:nftau}) from the version published in \cite{Kononovicius2013EPL,Gontis2014PlosOne} as the last numerical investigation confirms the possibility to include the term $x_f \varepsilon_{fc}$ into the denominator of first term. Source code of the numerical implementation, as well as compiled program and scripts used to produce data shown in figures may be found on GitHub (see \href{http://git.io/vJe5e}{http://git.io/vJe5e}).

Eqs. (\ref{eq:nftau})-(\ref{eq:taunfxi}) form a core of the consentaneous model of financial markets derived from the three state herding model \cite{Gontis2014PlosOne}. These equations belong to the class of non-linear stochastic differential equations exhibiting power law statistics and scaling of variable. Modeling of financial markets by these equations is comparable only with nonlinear GARCH(1,1) process \cite{Kononovicius2015PhysA}. As we have shown in previous work, such financial market model is able to reproduce power-law statistics of financial markets in very details and the most exciting aspect of this approach is that model parameters are the same for all markets from Vilnius to New York, and for all stocks. The return volatility in this model is defined by deviations of market price from its fundamental value $p(t)$ as a result of endogenous agent population dynamics in the set of three states. Here we consider the possibility to control market by introducing agents with predefined behavior.

\begin{figure}
\centering
\includegraphics[width=7cm]{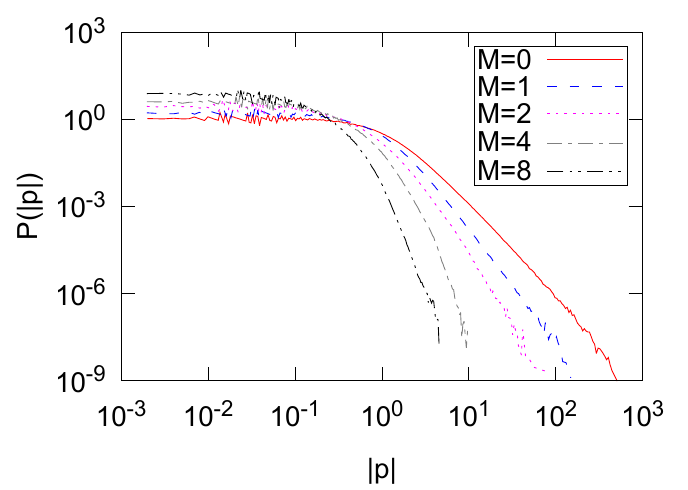}
\caption{Stationary PDF of absolute log-price, $|p(t)|$, when predefined fundamentalists, $M = 0;1;2;4;8$, are present. Results were obtained by numerically solving equations (\ref{eq:nftau})-(\ref{eq:taunfxi}). Parameters were as follows: $\tilde{\varepsilon}_{cf}=0.1+M$, $\varepsilon_{fc}=3$, $\varepsilon_{cc}=3$,  $H=300$, $a=0.5$, $\alpha=2$.}
\label{fig:fundam}
\end{figure}

It is obvious that the most effective market control method would be introduction of $M$ agents with predefined fundamentalist trading behavior. This would increase parameter $\sigma_{cf}$ by $h M$, or $\tilde{\varepsilon}_{cf} = \varepsilon_{cf} + M$. Observe, in Fig. \ref{fig:fundam}, that as $M$ increases the PDF of $p(t)$ becomes narrower - larger deviations of $p(t)$ become significantly less probable and standard deviation decreases. This process may be also seen as a convergence of $q$-Gaussian-like (power-law asymptotic behavior) distribution towards Gaussian-like distribution (exponential asymptotic behavior) - similarly to what was discussed in \cite{Kononovicius2014EPJB}. Evidently this control strategy yields excellent results, but its main drawback is a simple fact that currently there is no conventional agreement on how to estimate fundamental price, though some approaches are being conducted by performing behavioral experiments \cite{Hommes2006Elsevier,Hommes2006Tinbergen,Hommes2009Cheltenham}.

The beneficial role of noise in many physical systems is widely recognized. It is a well known that certain amounts of noise allows to strengthen the actual patterns exhibited by the dynamical system \cite{Mantegna1996PhysRevLett,GarciaOjalvo1999Springer,RouvasNicolis2007,Caruso2010PhysRevLett}. While the idea itself is not new to physics it was just only recently applied to socio-economic systems. E.g., it was shown that random promotions might lead to more efficient hierarchical structures \cite{Pluchino2010PhysA} as well as ``accidental'' politicians potentially improving legislature process \cite{Pluchino2011PhysA}. In a couple of more recent publications this idea was applied to the financial markets. In terms of the financial markets this idea appears to be somewhat controversial as the Efficient Market Hypothesis suggests that non-rational agents should be driven out from the market \cite{Friedman1956Princeton}. But the stochastic trading appears to work both as investment strategy as well as extreme event prevention strategy, at least in generic setups \cite{Biondo2013PhysRevE,Biondo2013PlosOne}. This kind of approach would be of great value as introduction of stochastic traders, unlike fundamentalist traders, is very simple in terms of implementation. Though the realistic introduction of the stochastic agents into the three state model, and their impact on the macroscopic behavior, is not straightforward.

The main problem is to define how the introduction of stochastic agents will impact the population dynamics between agent groups. Recall that Kirman�s herding model is an ad hoc Markov process on the microscopic, individual agent, level. Agents are free of any rationality, they are assumed to have zero intelligence. Namely they change the behavior with certain probability just in response to the contact with another agent. In the considered model this contact is equivalent to a market transaction.

Recall that there are two independent processes of agent population dynamics in proposed model (see Fig. \ref{fig:schema2}): fundamentalists-chartists and optimists-pessimists. As time scales of these processes are different up to three orders of magnitude we can assume them to be totally independent. In this approximation agents participate in both two state processes simultaneously.

In the Section \ref{sec:generic} we have already discussed the impact of stochastic agents on the population dynamics between two agent groups. Here we apply the same logic - stochastic agents influence ordinary agents to switch to the direct opposites of the considered groups. In the slow fundamentalist-chartist process, when a fundamentalist makes a trade with a stochastic agent, he perceives stochastic agent as chartist. While, on the other hand, when a chartist makes trade with a stochastic agent, the chartist perceives stochastic agent as fundamentalist. In a similar way for the faster optimist-pessimist process, optimists perceive stochastic agents as pessimists, while pessimists - as optimists. In all cases any ordinary agent can trade with only a half of stochastic agents, those who submit opposite trade orders. Therefore in both, fast and slow, processes stochastic agents should be perceived as an additional $M/2$ agents belonging to the direct opposite of any considered group.

\begin{figure}
\centering
\includegraphics[width=7cm]{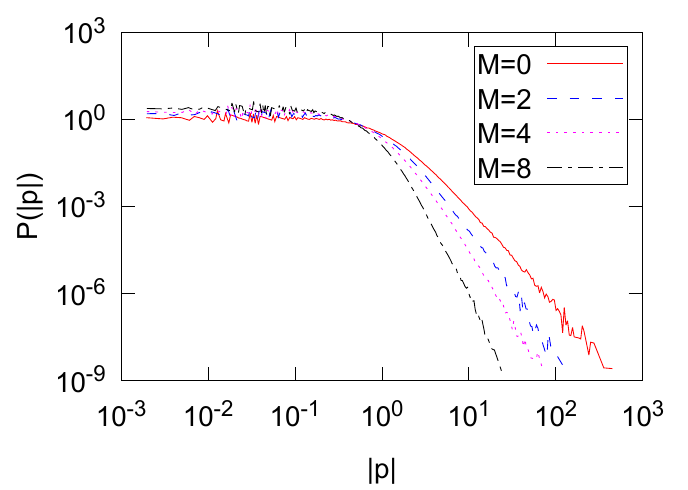}
\caption{Stationary PDF of absolute log-price, $|p(t)|$, in case when stochastic traders, $M = 0;2;4;8$, are assumed to have symmetric impact. Results were  obtained by numerically from equations (\ref{eq:nftau})-(\ref{eq:taunfxi}). Parameters were set as follows: $\tilde{\varepsilon}_{cc} = 3 + M/2$  $\tilde{\varepsilon}_{fc} = 3 + M/2$, $\tilde{\varepsilon}_{cf} = 0.1 + M/2$,  $H=300$, $a=0.5$, $\alpha=2$.}
\label{fig:stoch-all}
\end{figure}

Mathematically the impact of stochastic agents in this setup of two independent herding processes may be formalized in the following way: $\tilde{\varepsilon}_{fc} = \varepsilon_{fc} + M/2$, $\tilde{\varepsilon}_{cf} = \varepsilon_{cf} + M/2$, $\tilde{\varepsilon}_{op} = \varepsilon_{op} + M/2$ and $\tilde{\varepsilon}_{po} = \varepsilon_{po} + M/2$. In Fig. \ref{fig:stoch-all} we demonstrate that stochastic trading has a considerable effect diminishing price deviations from the fundamental value.

\section{Conclusions}
\label{sec:conclusions}

In this contribution we have considered control possibilities of the financial market providing our interpretation based on the artificial agent-based set-up built upon the three state herding model and reproducing main stylized facts of the markets. The three state model has its roots in generic Kirman's model, which provides a mathematical background for the wisdom of the crowd effect in social communities. Consequently this model reproduces extreme events, which can be seen as extreme deviations from the ``correct'' opinion, happening simply due to endogenous interactions. Fortunately, the core feature of the endogenous interactions causing extreme deviations, herding behavior, might be also used to prevent these extreme events. We propose to use predefined fundamentalist and stochastic traders, which through global herding coupling may be able to prevent extreme events in the financial markets.

First of all we demonstrate the efficiency of controlled agents in the simplest case. Namely we consider analytically the two state herding model with controlled agents present. Population dynamics driven by herding interactions can be manipulated by introducing controlled agents into the predefined states or just letting them change their state stochastically. From the theoretical point of view this simple case serves as an example of a stochastic system on the macroscopic level, which can be stabilized by introducing agents, who act stochastically.

Next we move on to the analysis of the market price deviations from its fundamental value in the three state herding model of financial markets, which is known to be capable of reproducing the empirical data \cite{Kononovicius2013EPL,Gontis2014PlosOne}. Numerical simulations using this model show that introduction of predefined fundamentalist traders is a very efficient method to reduce market price fluctuations. In this context we have also tested an extreme event prevention strategy (proposed in \cite{Biondo2013PhysRevE,Biondo2013PlosOne}), which assumes that stochastically trading agents might also stabilize financial market fluctuations. Introduction of the such agents into the three state model requires some interpretation how different agent groups would perceive stochastic traders. We do consider as the most reasonable approach that the  impact of stochastic traders on the dynamics of fundamentalists and chartists is symmetric. In this interpretation the macroscopic dynamics of market price falls under observable stabilization.


\end{document}